\documentclass[12pt,a4paper]{article}
\usepackage{amsmath,amsfonts,amsthm,graphicx,lscape}
\usepackage[dvips]{psfrag}
\usepackage{cite}

\usepackage{url}
\usepackage[normalem]{ulem}

\usepackage{textcomp}
\usepackage{amsmath,amsthm,amssymb,mathrsfs,cases}
\usepackage{amsfonts,graphicx,lscape}

\usepackage{accents}
\makeatletter
\def\widebar{\accentset{{\cc@style\underline{\mskip10mu}}}}
\makeatother

\numberwithin{equation}{section}

\def\beqa{\begin{eqnarray}}
\def\enqa{\end{eqnarray}}
\def\beq{\begin{equation}}
\def\enq{\end{equation}}

\begin{document}
\title{Integrable semi-discretizations of the sine-Gordon equation 
in non-characteristic
coordinates}
\author{Takayuki \textsc{Tsuchida}}
\maketitle
\begin{abstract} 
Integrable discretizations of the sine-Gordon equation in 
characteristic (or light-cone) 
coordinates have been extensively studied after 
the seminal 
works of Hirota and Orfanidis in the late 1970s. 
In contrast, integrable discretizations of the sine-Gordon equation 
in non-characteristic coordinates have been scarcely studied 
except the lattice sine-Gordon model proposed by Izergin and Korepin in 
the early 1980s. 
In this paper, using the 
zero-curvature 
representation, 
we 
propose 
integrable space discretizations of the sine-Gordon equation 
in three 
distinct 
cases of 
non-characteristic coordinates. 
For the most interesting case of the sine-Gordon equation 
in laboratory coordinates, 
the integrable space discretization is unwieldy; 
as a remedy, 
we rewrite the sine-Gordon equation 
as a 
two-component evolutionary system 
and present 
an aesthetically acceptable 
space discretization. 
\end{abstract}
%

\newpage
\noindent
\tableofcontents

\newpage
\section{Introduction}

Among a large number of integrable systems in 
\mbox{$1+1$} 
dimensions, 
the sine-Gordon equation 
stands out as 
one of the most 
important 
systems 
with wide 
applicability 
in mathematical physics~\cite{Scott71,Scott73,Rogers02}. 
With the increasing interest in discrete integrable systems in the late 1970s, 
Hirota~\cite{Hirota77} proposed an integrable discretization of the sine-Gordon equation, 
which is a fully discrete analog of the sine-Gordon equation in 
characteristic (or light-cone) coordinates. 
By 
developing Hirota's idea, 
Orfanidis~\cite{Orfa-78-1,Orfa-78-2} obtained 
integrable semi-discretizations 
(discretization of one of the two independent variables) 
of the sine-Gordon equation in 
characteristic coordinates. 
Hirota's work~\cite{Hirota77} and Orfanidis's work~\cite{Orfa-78-1,Orfa-78-2}
revealed 
a close relationship between 
auto-B\"acklund 
transformations and 
integrable discretizations for the particular example of the sine-Gordon 
equation in characteristic coordinates.  
In the continuous limit, Hirota's fully discrete sine-Gordon equation 
and Orfanidis's semi-discrete sine-Gordon equations reduce to the sine-Gordon equation in 
characteristic coordinates: 
\begin{equation}
\label{sGeq1}
u_{\xi \eta} = \sin u,
\end{equation}
where the subscripts denote 
partial differentiation. 

In the continuous space and time, the sine-Gordon equation in 
characteristic coordinates (\ref{sGeq1}) and 
the sine-Gordon equation in laboratory coordinates: 
\begin{equation}
\label{sGeq2}
u_{tt} - u_{xx} + \sin u =0,
\end{equation}
can be transformed into each other by a linear change of coordinates
and 
the integrability properties of the former 
can be obtained from the latter and vice versa~\cite{KN78-2,TF79}. 
However, this equivalence 
does not carry over 
directly 
to the 
discrete case 
and we cannot obtain 
an integrable semi-discretization of the sine-Gordon equation in laboratory coordinates 
(\ref{sGeq2}) 
from 
the results of Hirota~\cite{Hirota77} and Orfanidis~\cite{Orfa-78-1,Orfa-78-2}. 
%
Indeed, in the semi-discrete case, 
we cannot consider 
a natural change of coordinates 
that mixes 
the continuous independent variable 
and 
the discrete independent variable. 
In view of this,
we address a relatively less studied problem:\ 
how to construct integrable semi-discretizations of the sine-Gordon equation in 
non-characteristic coordinates, 
including 
the most interesting 
case of laboratory coordinates 
(\ref{sGeq2}). 
In 1981, 
Izergin and Korepin~\cite{IzeKor81}
proposed a quantum version of 
the lattice sine-Gordon model, 
which 
is an integrable 
space discretization of a two-component system 
equivalent to 
the sine-Gordon equation in laboratory coordinates 
(\ref{sGeq2}); 
a comprehensive description of 
the classical lattice sine-Gordon model 
can be found in~\cite{IzeKor86,Tarasov86,BogoIzeKor93}. 
However, it is not so easy to understand intuitively that 
the lattice sine-Gordon model 
reduces to the sine-Gordon equation in laboratory coordinates 
(\ref{sGeq2}) in the continuous limit. 

This paper is organized as follows. 
In section~2, 
we give 
the general description of 
the 
zero-curvature representation 
in the continuous 
and space-discrete cases, respectively. 
Then, we present 
the Lax pair for the continuous 
sine-Gordon equation in 
characteristic coordinates (\ref{sGeq1}) 
and rewrite it 
for 
non-characteristic coordinates in a way similar to~\cite{AKNS73-2}.  
In the subsequent sections, we 
mainly consider how to discretize 
the spatial part of 
the Lax pair. 
In section~3, we propose an integrable space discretization
(discretization of the spatial variable $x$) 
of the sine-Gordon equation in non-characteristic coordinates: 
\begin{equation}
\nonumber 
u_{xt} - u_{xx} = \sin u. 
\end{equation}
In section~4, we 
propose 
an integrable space discretization
of the sine-Gordon equation in non-characteristic coordinates: 
\begin{equation}
\nonumber 
u_{tt} + u_{xt} = \sin u. 
\end{equation}
In section~5, we rewrite the sine-Gordon equation in laboratory coordinates (\ref{sGeq2}) as 
a two-component evolutionary system: 
\begin{equation}
\nonumber
  \left\{ 
  \begin{array}{l} 
	u_t =v, \\
	v_t=u_{xx}-\sin u,
  \end{array} 
  \right.
\end{equation}
and present 
an integrable space discretization
of this system. 
In section~6, we 
obtain an integrable space discretization 
of the sine-Gordon equation in laboratory coordinates (\ref{sGeq2}) 
explicitly expressible as a scalar equation. 
Section~7 is devoted to conclusions. 
Throughout 
the paper, 
the sign in front of 
$\sin u$
can be changed 
by shifting the dependent variable $u$ as 
\mbox{$u \to u \pm \pi$}.

\section{Zero-curvature 
representation}

\subsection
{General description} 
\label{section2.1}

In the continuous space and time, we 
consider the pair of 
linear equations~\cite{AKNS74,ZS79}: 
\begin{subequations}
\label{line}
\begin{align}
& \Psi_{x} = U (\lambda)\Psi, 
\label{line_s}
\\
& \Psi_{t} = V (\lambda)\Psi.
\label{line_t}
\end{align}
\end{subequations}
Here, 
the subscripts denote 
partial differentiation, 
$\Psi$ is a 
column vector 
and the square 
matrices $U (\lambda)$
and $V (\lambda)$ 
constitute the Lax pair~\cite{Lax} depending on the constant spectral parameter $\lambda$. 
The compatibility condition of the overdetermined linear equations 
(\ref{line}) 
is given by the zero-curvature equation~\cite{AKNS74,ZS79}: 
\begin{equation}
 U_t-V_x+UV-VU=0. 
\label{Lax_eq00}
\end{equation}

As a space-discrete analog of 
(\ref{line}), 
we consider the pair of semi-discrete 
linear equations~\cite{AL1,AL76}: 
\begin{subequations}
\label{sd_line}
\begin{align}
& \Psi_{n+1} = L_n (\lambda)
\Psi_n, 
\label{line_s2}
\\
& \Psi_{n,t} = V_n (\lambda)\Psi_n.
\label{line_t2}
\end{align}
\end{subequations}
Here, \mbox{$n \in \mathbb{Z}$} is the discrete spatial variable, 
the subscript $t$ denotes the 
time derivative and $\lambda$ is the constant 
spectral parameter; 
$\Psi_n$ is a 
column vector 
and the square 
matrices $L_n$
and $V_n$ 
constitute the 
Lax pair~\cite{Lax}. 
The linear equations (\ref{sd_line}) 
reduce to (\ref{line}) in the continuous 
limit 
\mbox{$\varDelta \to 0$}
by 
assuming the asymptotic expansions: 
\begin{equation}
L_n (\lambda) = I + \varDelta \hspace{1pt} U (\lambda) + \mathcal{O}(\varDelta^2), \hspace{5mm} 
V_n (\lambda) = V(\lambda) + \mathcal{O}(\varDelta), 
\nonumber 
\end{equation}
where $I$ is the identity matrix and $\varDelta$ is a 
lattice parameter. 

The compatibility condition of the overdetermined linear equations 
(\ref{sd_line}) 
is given by the 
semi-discrete zero-curvature equation~\cite{
Kako,Levi80,Ize81}: 
\begin{equation}
 L_{n,t} = V_{n+1}L_n - L_n V_n, 
\label{sdLax_eq00}
\end{equation}
%
which 
implies the 
relation: 
\begin{equation}
\frac{\partial}{\partial t} \log (\det L_n) 
= 
\mathrm{tr} \hspace{1pt} V_{n+1} - \mathrm{tr} \hspace{1pt} V_{n}. 
\label{sd-cons}
\end{equation}

In this paper, we 
focus on the case of a traceless matrix $V_n$, 
so we require that the determinant of 
$L_n$ should be a constant. 
Moreover, we also 
assume that 
$V_n (\lambda)$ in the semi-discrete case 
and $V (\lambda)$ in the continuous case 
have 
the same 
functional 
form 
with respect to $\lambda$. 

\subsection
{The sine-Gordon equation in characteristic coordinates} 
\label{section2.2}

We first consider the 
following 
\mbox{$2 \times 2$} Lax pair~\cite{AKNS73-2}:  
\begin{subequations}
\label{AKNS_Lax}
\begin{align}
& U (\lambda)= 
\lambda \left[
\begin{array}{cc}
 -1 & 0 \\
 0 & 1 \\
\end{array}
\right] + \left[
\begin{array}{cc}
 0 & -\frac{u_x}{2} \\
 \frac{u_x}{2} & 0 \\
\end{array}
\right],
\label{}
\\[2mm]
& V (\lambda)
= 
-\frac{1}{4\lambda} \left[
\begin{array}{cc}
 \cos u & \sin u \\
 \sin u & -\cos u \\
\end{array}
\right].
\label{}
\end{align}
\end{subequations}
By setting \mbox{$\lambda= \mathrm{i}\zeta$} where $\mathrm{i}$ is the imaginary unit, 
(\ref{AKNS_Lax}) coincides with 
the original Lax pair proposed 
by 
Ablowitz, Kaup, Newell and Segur~\cite{AKNS73-2}. 
Substituting the Lax pair (\ref{AKNS_Lax}) into the zero-curvature equation 
(\ref{Lax_eq00}),  
we obtain the sine-Gordon equation in 
characteristic coordinates: 
\begin{equation}
\nonumber
u_{xt} = \sin u.
\end{equation}

\subsection
{The sine-Gordon equation in non-characteristic coordinates} 
\label{section2.3}

As a generalization of 
(\ref{AKNS_Lax}) (cf.~\cite{AKNS73-2}), 
we consider the following Lax pair: 
\begin{subequations}
\label{TF_Lax}
\begin{align}
& U (\lambda)= a\left\{ 
\lambda \left[
\begin{array}{cc}
 -1 & 0 \\
 0 & 1 \\
\end{array}
\right] + \left[
\begin{array}{cc}
 0 & -\frac{c u_t + d u_x}{2(ad+bc)} \\
 \frac{c u_t + d u_x}{2(ad+bc)} & 0 \\
\end{array}
\right] \right\} + 
\frac{c}{\lambda} \left[
\begin{array}{cc}
 \cos u & \sin u \\
 \sin u & -\cos u \\
\end{array}
\right],
\label{}
\\[2mm]
& V (\lambda)
= b\left\{ 
\lambda \left[
\begin{array}{cc}
 -1 & 0 \\
 0 & 1 \\
\end{array}
\right] + \left[
\begin{array}{cc}
 0 & -\frac{c u_t + d u_x}{2(ad+bc)} \\
 \frac{c u_t + d u_x}{2(ad+bc)} & 0 \\
\end{array}
\right] \right\} 
-\frac{d}{\lambda} \left[
\begin{array}{cc}
 \cos u & \sin u \\
 \sin u & -\cos u \\
\end{array}
\right].
\label{}
\end{align}
\end{subequations}
Here, $a$, $b$, $c$ and $d$ are constants satisfying the conditions 
\mbox{$(a,c) \neq (0,0)$}, \mbox{$(b,d) \neq (0,0)$} and \mbox{$ad+bc \neq 0$}. 
Substituting the Lax pair (\ref{TF_Lax}) into the zero-curvature equation 
(\ref{Lax_eq00}),  
we obtain the sine-Gordon equation 
in the parametric form: 
\begin{equation}
\label{sG_nonlight}
\left( a \partial_t - b \partial_x \right) \left( c \partial_t + d \partial_x \right) u
 = 4 (ad+bc)^2 \sin u.
\end{equation}

In the subsequent sections, we focus on 
the following three cases of non-characteristic coordinates:\  
(i) \mbox{$a=0$},  
(ii) 
\mbox{$d=0$} and 
(iii) \mbox{$ad=bc$}, 
which 
result in 
the disappearance of 
(i) $u_{tt}$, (ii) $u_{xx}$ and (iii) $u_{xt}$ 
from (\ref{sG_nonlight}), respectively.

\section{Discretization 
of the sine-Gordon equation \mbox{$u_{xt} - u_{xx} = \sin u$}}

In the case \mbox{$a=0$} (so \mbox{$bc \neq 0$}), 
the sine-Gordon equation 
(\ref{sG_nonlight}) reads 
\begin{equation}
\label{sG_3.1}
 u_{xt} +\frac{d}{c} u_{xx} = -4 bc \sin u.
\end{equation}
By setting \mbox{$b=-\frac{1}{4c}$} and \mbox{$d=-c$}, 
(\ref{sG_3.1}) reduces to the canonical form: 
\begin{equation}
\label{sG_3.2}
 u_{xt} -u_{xx} = \sin u. 
\end{equation}
The Lax pair for (\ref{sG_3.2}) is given by 
\begin{subequations}
\label{TF_Lax2}
\begin{align}
& U (\lambda)= 
\frac{c}{\lambda} \left[
\begin{array}{cc}
 \cos u & \sin u \\
 \sin u & -\cos u \\
\end{array}
\right],
\label{}
\\[2mm]
& V (\lambda)
= -\frac{\lambda}{4c} \left[
\begin{array}{cc}
 -1 & 0 \\
 0 & 1 \\
\end{array}
\right] + \left[
\begin{array}{cc}
 0 & -\frac{u_t -u_x}{2} \\
 \frac{u_t -u_x}{2} & 0 \\
\end{array}
\right] 
+\frac{c}{\lambda} \left[
\begin{array}{cc}
 \cos u & \sin u \\
 \sin u & -\cos u \\
\end{array}
\right].
\label{}
\end{align}
\end{subequations}
Note that we can set \mbox{$c=1$} in (\ref{TF_Lax2}) by rescaling the spectral parameter $\lambda$. 

We first 
consider the 
following 
semi-discrete 
Lax pair: 
\begin{subequations}
\label{sd_Lax1}
\begin{align}
& L_n (\lambda) = \left[
\begin{array}{cc}
 1 & 0 \\
 0 & 1 \\
\end{array}
\right] + 
\frac{\varDelta}{\lambda} \left[
\begin{array}{cc}
 \cos \left(\frac{u_{n+1}+u_n}{2} \right) & \sin \left( \frac{u_{n+1}+u_n}{2} \right) \\[1mm]
 \sin \left( \frac{u_{n+1}+u_n}{2} \right) & -\cos \left( \frac{u_{n+1}+u_n}{2} \right) \\
\end{array}
\right],
\label{}
\label{L_n_Orf}
\\[2mm]
& V_n (\lambda)
= b\lambda \left[
\begin{array}{cc}
 -1 & 0 \\
 0 & 1 \\
\end{array}
\right] + \left[
\begin{array}{cc}
 0 & f_n \\
 -f_n  & 0 \\
\end{array}
\right] 
-\frac{d}{\lambda} \left[
\begin{array}{cc}
 \cos u_n & \sin u_n \\
 \sin u_n & -\cos u_n \\
\end{array}
\right],
\label{}
\\[2mm]
& f_n:=\varDelta \hspace{1pt} b \hspace{1pt} \sin\left( \frac{u_n+u_{n-1}}{2} \right) - \frac{u_{n,t}+u_{n-1,t}}{4}
	-\frac{d}{\varDelta} \sin\left( \frac{u_n-u_{n-1}}{2} \right), 
\end{align}
\end{subequations}
where 
$\varDelta$ is a (typically small but nonzero) lattice parameter. 
A discrete 
spatial Lax matrix similar to (\ref{L_n_Orf}) 
appeared in \cite{Levi80}. 
Substituting the Lax pair (\ref{sd_Lax1}) into the semi-discrete zero-curvature equation (\ref{sdLax_eq00}), 
we obtain 
\begin{align}
&  u_{n+1,t}-u_{n-1,t} + \frac{4d}{\varDelta} \left[ \sin\left( \frac{u_{n+1}-u_n}{2} \right)- \sin\left( \frac{u_n-u_{n-1}}{2} \right) \right]
\nonumber \\[1mm]
& = -4\varDelta \hspace{1pt} b \left[ \sin\left( \frac{u_{n+1}+u_n}{2} \right) + \sin\left( \frac{u_n+u_{n-1}}{2} \right) \right]. 
\label{sdsG1}
\end{align}
This 
can be considered a space-discrete analog of (\ref{sG_3.1}). 
By further setting \mbox{$b=-\frac{1}{4}$} and \mbox{$d=-1$}, 
(\ref{sdsG1}) reduces to 
\begin{align}
&  \frac{u_{n+1,t}-u_{n-1,t}}{2\varDelta} 
	- \frac{2}{\varDelta^2} \left[ \sin\left( \frac{u_{n+1}-u_n}{2} \right)- \sin\left( \frac{u_n-u_{n-1}}{2} \right) \right]
\nonumber \\[1mm] 
& = \frac{1}{2} \left[ \sin\left( \frac{u_{n+1}+u_n}{2} \right) + \sin\left( \frac{u_n+u_{n-1}}{2} \right) \right], 
\label{sdsG2}
\end{align}
which is an integrable space discretization of the sine-Gordon equation 
in non-characteristic coordinates (\ref{sG_3.2}). 

Alternatively, we 
can also consider the 
semi-discrete Lax pair: 
\begin{subequations}
\label{sd_Lax2}
\begin{align}
& L_n (\lambda) = \lambda \left[
\begin{array}{cc}
 -\varDelta \hspace{1pt} a & 0 \\
 0 & \varDelta \hspace{1pt} a \\
\end{array}
\right] + \left[
\begin{array}{cc}
 \cos \left(\frac{u_{n+1}-u_n}{2} \right) & -\sin \left( \frac{u_{n+1}-u_n}{2} \right) \\[1mm]
 \sin \left( \frac{u_{n+1}-u_n}{2} \right) & \cos \left( \frac{u_{n+1}-u_n}{2} \right) \\
\end{array}
\right],
\label{}
\\[2mm]
& V_n (\lambda)
= b\lambda \left[
\begin{array}{cc}
 -1 & 0 \\
 0 & 1 \\
\end{array}
\right] + \left[
\begin{array}{cc}
 0 & g_n \\
 -g_n  & 0 \\
\end{array}
\right] 
-\frac{d}{\lambda} \left[
\begin{array}{cc}
 \cos u_n & \sin u_n \\
 \sin u_n & -\cos u_n \\
\end{array}
\right],
\label{}
\\[2mm]
& g_n:=\varDelta \hspace{1pt} ad \hspace{1pt} \sin\left( \frac{u_n+u_{n-1}}{2} \right) - \frac{u_{n,t}-u_{n-1,t}}{4}
	-\frac{b}{\varDelta\hspace{1pt}a} \sin\left( \frac{u_n-u_{n-1}}{2} \right). 
\end{align}
\end{subequations}
Substituting the Lax pair (\ref{sd_Lax2}) into the semi-discrete zero-curvature equation (\ref{sdLax_eq00}), 
we obtain 
\begin{align}
&  u_{n+1,t}-u_{n-1,t} 
	- \frac{4b}{\varDelta\hspace{1pt}a} \left[ \sin\left( \frac{u_{n+1}-u_n}{2} \right)- \sin\left( \frac{u_n-u_{n-1}}{2} \right) \right]
\nonumber \\[1mm] 
& = 4\varDelta \hspace{1pt} ad \left[ \sin\left( \frac{u_{n+1}+u_n}{2} \right) + \sin\left( \frac{u_n+u_{n-1}}{2} \right) \right]. 
\label{sdsG3}
\end{align}
By further setting \mbox{$b=a$} and \mbox{$d=\frac{1}{4a}$}, 
(\ref{sdsG3}) reduces to 
\begin{align}
&  \frac{u_{n+1,t}-u_{n-1,t}}{2\varDelta} 
	- \frac{2}{\varDelta^2} \left[ \sin\left( \frac{u_{n+1}-u_n}{2} \right)- \sin\left( \frac{u_n-u_{n-1}}{2} \right) \right]
\nonumber \\[1mm] 
& = \frac{1}{2} \left[ \sin\left( \frac{u_{n+1}+u_n}{2} \right) + \sin\left( \frac{u_n+u_{n-1}}{2} \right) \right], 
\nonumber
\end{align}
which coincides with (\ref{sdsG2}). 
Thus, we have two different 
Lax pairs 
for 
the integrable space discretization 
(\ref{sdsG2}) 
of the sine-Gordon equation 
in non-characteristic coordinates (\ref{sG_3.2}).

\section{Discretization 
of the sine-Gordon equation \mbox{$u_{tt} + u_{xt} = \sin u$}}

In the case \mbox{$d=0$} (so \mbox{$bc \neq 0$}), 
the sine-Gordon equation 
(\ref{sG_nonlight}) reads 
\begin{equation}
\label{sG_4.1}
 \frac{a}{b} u_{tt} -u_{xt} = 4 bc \sin u.
\end{equation}
By setting \mbox{$a=-b$} and \mbox{$c=-\frac{1}{4b}$}, 
(\ref{sG_4.1}) reduces to the canonical form: 
\begin{equation}
\label{sG_4.2}
 u_{tt} +u_{xt} = \sin u. 
\end{equation}
The Lax pair for (\ref{sG_4.2}) is given by 
\begin{subequations}
\label{TF_Lax3}
\begin{align}
& U (\lambda)= b\lambda \left[
\begin{array}{cc}
 1 & 0 \\
 0 & -1 \\
\end{array}
\right] + \left[
\begin{array}{cc}
 0 & \frac{u_t}{2} \\
 -\frac{u_t}{2} & 0 \\
\end{array}
\right] - 
\frac{1}{4b\lambda} \left[
\begin{array}{cc}
 \cos u & \sin u \\
 \sin u & -\cos u \\
\end{array}
\right],
\label{}
\\[2mm]
& V (\lambda)
= b \lambda \left[
\begin{array}{cc}
 -1 & 0 \\
 0 & 1 \\
\end{array}
\right] + \left[
\begin{array}{cc}
 0 & -\frac{u_t}{2} \\
 \frac{u_t}{2} & 0 \\
\end{array}
\right].
\label{}
\end{align}
\end{subequations}
Note that we can set \mbox{$b=1$} in (\ref{TF_Lax3}) by rescaling the spectral parameter $\lambda$. 

We consider the following semi-discrete Lax pair: 
\begin{subequations}
\label{sd_Lax3}
\begin{align}
& L_n (\lambda) = \left[
\begin{array}{cc}
 1 & 0 \\
 0 & 1 \\
\end{array}
\right] + 
\varDelta \left[
\begin{array}{cc}
 \lambda + X_n - \frac{1}{4\lambda} \cos u_n & -\frac{h_{n+1} + h_{n}}{2} - \frac{1}{4\lambda} \sin u_n \\[1mm]
 \frac{h_{n+1} + h_{n}}{2} - \frac{1}{4\lambda} \sin u_n & -\lambda + X_n + \frac{1}{4\lambda} \cos u_n\\
\end{array}
\right],
\label{}
\label{L_n_Orf3}
\\[2mm]
& V_n (\lambda)
= \lambda \left[
\begin{array}{cc}
 -1 & 0 \\
 0 & 1 \\
\end{array}
\right] + \left[
\begin{array}{cc}
 0 & h_n \\
 -h_n  & 0 \\
\end{array}
\right],
\label{}
\end{align}
\end{subequations}
where 
$\varDelta$ is a (typically small but nonzero) lattice parameter 
and 
$X_n$ and $h_n$ are some functions of the 
dependent variable $u_n$ 
to be determined. 

Substituting the Lax pair (\ref{sd_Lax3}) into the semi-discrete zero-curvature equation (\ref{sdLax_eq00}), 
we obtain 
\begin{subequations}
\label{system_4.5}
\begin{align}
& u_{n,t} + h_{n+1}+ h_{n}=0,
\label{4.5.1}
\\[1mm]
& X_{n,t} - \frac{h_{n+1}^2 -h_n^2}{2}
=0,
\label{4.5.2}
\\[1mm]
&\left( h_{n+1}+h_{n} \right)_t + \sin u_n +\frac{2}{\varDelta} \left( 1+ \varDelta X_n \right)
	\left( h_{n+1}-h_{n} \right)=0.  
\label{4.5.3}
\end{align}
\end{subequations}
The system (\ref{system_4.5}) 
implies the relation:
\begin{equation}
\left[ \frac{\varDelta^2}{4} \left( h_{n+1}+h_{n} \right)^2 + \left( 1+ \varDelta X_n \right)^2 + \frac{\varDelta^2}{2} \cos u_n 
\right]_t = 0,
\nonumber
\end{equation}
which
can also be 
obtained as a direct consequence of 
(\ref{sd-cons}). 
Thus, we set 
\begin{equation}
\frac{\varDelta^2}{4} \left( u_{n,t} \right)^2 
	+ \left( 1+ \varDelta X_n \right)^2 + \frac{\varDelta^2}{2} \cos u_n = 1, 
\nonumber
\end{equation}
and choose the solution of this quadratic equation for $1+\varDelta X_n$ as 
\begin{equation}
 1+ \varDelta X_n = \sqrt{1-\frac{\varDelta^2}{4} \left( u_{n,t} \right)^2 - \frac{\varDelta^2}{2} \cos u_n}, 
\label{X_n}
\end{equation}
where the square root function 
is defined as the Maclaurin series in $\varDelta^2$. 

From (\ref{4.5.1}), (\ref{4.5.3}) and (\ref{X_n}), we have two relations: 
\begin{align}
& h_n = -\frac{u_{n,t}}{2} -\frac{\frac{\varDelta}{4}\left( u_{n,tt}-\sin u_n \right)}
	{\sqrt{1-\frac{\varDelta^2}{4} \left( u_{n,t} \right)^2 - \frac{\varDelta^2}{2} \cos u_n}},
\nonumber
\\[1mm]
& h_{n+1} = -\frac{u_{n,t}}{2} +\frac{\frac{\varDelta}{4}\left( u_{n,tt}-\sin u_n \right)}
	{\sqrt{1-\frac{\varDelta^2}{4} \left( u_{n,t} \right)^2 - \frac{\varDelta^2}{2} \cos u_n}}, 
\nonumber
\end{align}
which together provide an integrable space discretization of the sine-Gordon equation 
in non-characteristic coordinates (\ref{sG_4.2}): 
\begin{equation}
 \frac{u_{n+1,tt}-\sin u_{n+1}}
	{\sqrt{1-\frac{\varDelta^2}{4} \left( u_{n+1,t} \right)^2 - \frac{\varDelta^2}{2} \cos u_{n+1}}}
+\frac{u_{n,tt}-\sin u_n}
	{\sqrt{1-\frac{\varDelta^2}{4} \left( u_{n,t} \right)^2 - \frac{\varDelta^2}{2} \cos u_n}}
 +\frac{2}{\varDelta}\left( u_{n+1,t} -u_{n,t} \right)  
=0.
\label{4.8}
\end{equation}
The explicit form of the semi-discrete Lax pair 
for 
(\ref{4.8}) is given by 
\begin{align}
L_n (\lambda) &= \sqrt{1-\frac{\varDelta^2}{4} \left( u_{n,t} \right)^2 - \frac{\varDelta^2}{2} \cos u_n}
\left[
\begin{array}{cc}
 1 & 0 \\
 0 & 1 \\
\end{array}
\right] 
\nonumber \\[1mm]
& \hphantom{=} \; + 
\varDelta \left[
\begin{array}{cc}
 \lambda - \frac{1}{4\lambda} \cos u_n & \frac{u_{n,t}}{2} - \frac{1}{4\lambda} \sin u_n \\[1mm]
 -\frac{u_{n,t}}{2} - \frac{1}{4\lambda} \sin u_n & -\lambda + \frac{1}{4\lambda} \cos u_n\\
\end{array}
\right],
\nonumber 
\\[3mm]
V_n (\lambda)
&= \lambda \left[
\begin{array}{cc}
 -1 & 0 \\
 0 & 1 \\
\end{array}
\right] + \left\{ \frac{u_{n,t}}{2} +\frac{\frac{\varDelta}{4}\left( u_{n,tt}-\sin u_n \right)}
	{\sqrt{1-\frac{\varDelta^2}{4} \left( u_{n,t} \right)^2 - \frac{\varDelta^2}{2} \cos u_n}}\right\}
\left[
\begin{array}{cc}
 0 & -1 \\
 1 & 0 \\
\end{array}
\right].
\nonumber
\end{align}

\section{Discretization 
of the 
system 
\mbox{$u_t=v, \;\; v_t= u_{xx}-\sin u$}}

In the case \mbox{$ad=bc (\neq 0)$}, the sine-Gordon equation (\ref{sG_nonlight}) reads 
\begin{equation}
\label{sG_nonlight4}
\frac{a^2}{b^2} u_{tt} - u_{xx} = 16 ac \sin u.
\end{equation}
By setting \mbox{$a=b$} and \mbox{$c=d=-\frac{1}{16a}$}, (\ref{sG_nonlight4}) reduces to 
the canonical form: 
\begin{equation}
\label{sG_light2}
u_{tt} - u_{xx} +\sin u=0.
\end{equation}
The Lax pair for (\ref{sG_light2}) is given by 
(cf.~\cite{AKNS73-2,TF74,ZTF75}) 
\begin{subequations}
\label{TF_Lax4}
\begin{align}
& U (\lambda)= a\lambda 
\left[
\begin{array}{cc}
 -1 & 0 \\
 0 & 1 \\
\end{array}
\right] + \left[
\begin{array}{cc}
 0 & -\frac{u_t + u_x}{4} \\
 \frac{u_t + u_x}{4} & 0 \\
\end{array}
\right] - 
\frac{1}{16a\lambda} \left[
\begin{array}{cc}
 \cos u & \sin u \\
 \sin u & -\cos u \\
\end{array}
\right],
\label{}
\\[2mm]
& V (\lambda)
= a \lambda 
\left[
\begin{array}{cc}
 -1 & 0 \\
 0 & 1 \\
\end{array}
\right] + \left[
\begin{array}{cc}
 0 & -\frac{u_t + u_x}{4} \\
 \frac{u_t + u_x}{4} & 0 \\
\end{array}
\right] 
+\frac{1}{16a\lambda} \left[
\begin{array}{cc}
 \cos u & \sin u \\
 \sin u & -\cos u \\
\end{array}
\right].
\label{}
\end{align}
\end{subequations}
Note that we can 
set \mbox{$a=\frac{1}{16}$} 
in (\ref{TF_Lax4}) by rescaling the spectral parameter $\lambda$. 

The sine-Gordon equation in laboratory coordinates (\ref{sG_light2}) can be rewritten as 
a two-component evolutionary system: 
\begin{equation}
\label{sG_2c}
  \left\{ 
  \begin{array}{l} 
	u_t =v, \\
	v_t=u_{xx}-\sin u,
  \end{array} 
  \right.
\end{equation}
for which the Lax pair 
is given by 
\begin{align}
& U (\lambda)= \frac{\lambda}{16} 
\left[
\begin{array}{cc}
 -1 & 0 \\
 0 & 1 \\
\end{array}
\right] + \left[
\begin{array}{cc}
 0 & -\frac{v + u_x}{4} \\
 \frac{v + u_x}{4} & 0 \\
\end{array}
\right] - 
\frac{1}{\lambda} \left[
\begin{array}{cc}
 \cos u & \sin u \\
 \sin u & -\cos u \\
\end{array}
\right],
\nonumber 
\\[2mm]
& V (\lambda)
= \frac{\lambda}{16} 
\left[
\begin{array}{cc}
 -1 & 0 \\
 0 & 1 \\
\end{array}
\right] + \left[
\begin{array}{cc}
 0 & -\frac{v + u_x}{4} \\
 \frac{v + u_x}{4} & 0 \\
\end{array}
\right] 
+\frac{1}{\lambda} \left[
\begin{array}{cc}
 \cos u & \sin u \\
 \sin u & -\cos u \\
\end{array}
\right].
\nonumber
\end{align}

To obtain 
an integrable space discretization
of the system (\ref{sG_2c}), 
we consider 
a generalization of 
the overdetermined 
linear equations (\ref{sd_line})
such that the 
functional 
form 
of the spatial Lax matrix $L_n$ 
can differ 
depending 
on the evenness/oddness of the lattice
site: 
\begin{subequations}
\label{sd_line2}
\begin{align}
& \Psi_{2n} = L_{2n-1} (\lambda) \Psi_{2n-1}, \hspace{5mm}  \Psi_{2n+1} = L_{2n} (\lambda) \Psi_{2n}, 
\label{line_s3}
\\[1mm] 
& \Psi_{2n-1,t} = V_{2n-1} (\lambda)\Psi_{2n-1}, \hspace{5mm} \Psi_{2n,t} = V_{2n} (\lambda)\Psi_{2n}. 
\label{line_t3}
\end{align}
\end{subequations}
Here,
$L_{2n-1} (\lambda)$ 
and 
$L_{2n} (\lambda)$ 
can have different 
dependence on the spectral parameter $\lambda$, 
whereas $V_{2n-1} (\lambda)$ 
and $V_{2n} (\lambda)$ have the same dependence on 
$\lambda$. 
The compatibility conditions of the overdetermined linear equations 
(\ref{sd_line2}) 
are given by the 
semi-discrete zero-curvature equations: 
\begin{subequations}
\label{sdLax_eq002} 
\begin{align}
& L_{2n-1,t} = V_{2n}L_{2n-1} - L_{2n-1} V_{2n-1}, 
\label{sdLax_eq002_1} \\[1mm]
& L_{2n,t} = V_{2n+1}L_{2n} - L_{2n} V_{2n}. 
\label{sdLax_eq002_2}
\end{align}
\end{subequations}

Specifically, 
as a combination of (\ref{sd_Lax1}) and (\ref{sd_Lax2}) (with the sign inversion of $\varDelta$), 
we  
consider the 
semi-discrete 
Lax pair
of the following form:
\begin{subequations}
\label{sd_Lax4}
\begin{align}
& L_{2n-1} (\lambda) = \left[
\begin{array}{cc}
 1 & 0 \\
 0 & 1 \\
\end{array}
\right] -
\frac{\varDelta}{\lambda} \left[
\begin{array}{cc}
 \cos \left(\frac{u_{2n}+u_{2n-1}}{2} \right) & \sin \left( \frac{u_{2n}+u_{2n-1}}{2} \right) \\[1mm]
 \sin \left( \frac{u_{2n}+u_{2n-1}}{2} \right) & -\cos \left( \frac{u_{2n}+u_{2n-1}}{2} \right) \\
\end{array}
\right],
\label{}
\label{L_n_Orf4}
\\[2mm]
& V_{2n-1} (\lambda)
= b\lambda \left[
\begin{array}{cc}
 -1 & 0 \\
 0 & 1 \\
\end{array}
\right] + \left[
\begin{array}{cc}
 0 & f_{2n-1} \\
 -f_{2n-1}  & 0 \\
\end{array}
\right] 
-\frac{d}{\lambda} \left[
\begin{array}{cc}
 \cos u_{2n-1} & \sin u_{2n-1} \\
 \sin u_{2n-1} & -\cos u_{2n-1} \\
\end{array}
\right],
\label{} 
\\[2mm]
& L_{2n} (\lambda) = \lambda \left[
\begin{array}{cc}
 \varDelta \hspace{1pt} a & 0 \\
 0 & -\varDelta \hspace{1pt} a \\
\end{array}
\right] + \left[
\begin{array}{cc}
 \cos \left(\frac{u_{2n+1}-u_{2n}}{2} \right) & -\sin \left( \frac{u_{2n+1}-u_{2n}}{2} \right) \\[1mm]
 \sin \left( \frac{u_{2n+1}-u_{2n}}{2} \right) & \cos \left( \frac{u_{2n+1}-u_{2n}}{2} \right) \\
\end{array}
\right],
\label{}
\label{L_n_Orf5}
\\[2mm]
& V_{2n} (\lambda)
= b\lambda \left[
\begin{array}{cc}
 -1 & 0 \\
 0 & 1 \\
\end{array}
\right] + \left[
\begin{array}{cc}
 0 & f_{2n} \\
 -f_{2n}  & 0 \\
\end{array}
\right] 
-\frac{d}{\lambda} \left[
\begin{array}{cc}
 \cos u_{2n} & \sin u_{2n} \\
 \sin u_{2n} & -\cos u_{2n} \\
\end{array}
\right],
\label{}
\end{align}
\end{subequations}
where 
$\varDelta$ is a (typically small but nonzero) lattice parameter 
and $f_{2n-1}$ and $f_{2n}$ are some functions to be determined. 

Substituting the Lax pair (\ref{sd_Lax4}) into the semi-discrete zero-curvature equations 
(\ref{sdLax_eq002_1}) and (\ref{sdLax_eq002_2}), 
we obtain 
\begin{subequations}
\label{system_5.9}
\begin{align}
& f_{2n}-f_{2n-1}+2\varDelta \hspace{1pt}b  \sin \left( \frac{u_{2n}+u_{2n-1}}{2} \right)=0,
\label{5.9.1}
\\[1mm]
& \varDelta \frac{u_{2n,t}+u_{2n-1,t}}{2} - 2d  \sin \left( \frac{u_{2n}-u_{2n-1}}{2} \right) + \varDelta (f_{2n}+f_{2n-1}) 
=0, 
\label{5.9.2}
\end{align}
\end{subequations} 
and 
\begin{subequations}
\label{system_5.10}
\begin{align}
& 2b \sin \left( \frac{u_{2n+1}-u_{2n}}{2} \right) - \varDelta \hspace{1pt}a (f_{2n+1}+f_{2n})=0,
\label{5.10.1}
\\[1mm]
& \frac{u_{2n+1,t}-u_{2n,t}}{2} +2\varDelta \hspace{1pt}ad \sin \left( \frac{u_{2n+1}+u_{2n}}{2} \right) + f_{2n+1}-f_{2n} =0, 
\label{5.10.2}
\end{align}
\end{subequations} 
respectively. 
Note that (\ref{system_5.9}) can be rewritten as 
\begin{subequations}
\label{system_5.11}
\begin{align}
& f_{2n} + \frac{u_{2n,t}+u_{2n-1,t}}{4} - \frac{d}{\varDelta}  \sin \left( \frac{u_{2n}-u_{2n-1}}{2} \right) 
	+\varDelta \hspace{1pt}b  \sin \left( \frac{u_{2n}+u_{2n-1}}{2} \right)=0,
\label{5.11.1}
\\[1mm]
& f_{2n-1} + \frac{u_{2n,t}+u_{2n-1,t}}{4} - \frac{d}{\varDelta}  \sin \left( \frac{u_{2n}-u_{2n-1}}{2} \right) 
	-\varDelta \hspace{1pt}b  \sin \left( \frac{u_{2n}+u_{2n-1}}{2} \right)=0,
\label{5.11.2}
\end{align}
\end{subequations} 
while 
(\ref{system_5.10}) can be rewritten as 
\begin{subequations}
\label{system_5.12}
\begin{align}
& f_{2n+1} + \frac{u_{2n+1,t}-u_{2n,t}}{4} - \frac{b}{\varDelta\hspace{1pt}a} \sin \left( \frac{u_{2n+1}-u_{2n}}{2} \right) 
	+\varDelta \hspace{1pt}ad \sin \left( \frac{u_{2n+1}+u_{2n}}{2} \right)=0,
\label{5.12.1}
\\[1mm]
& f_{2n} - \frac{u_{2n+1,t}-u_{2n,t}}{4} - \frac{b}{\varDelta\hspace{1pt}a} \sin \left( \frac{u_{2n+1}-u_{2n}}{2} \right) 
	-\varDelta \hspace{1pt}ad \sin \left( \frac{u_{2n+1}+u_{2n}}{2} \right)=0.
\label{5.12.2}
\end{align}
\end{subequations} 
By eliminating $f_{2n}$ from (\ref{5.11.1}) and (\ref{5.12.2}), we have  
\begin{subequations}
\label{5.13}
\begin{align}
& \frac{u_{2n+1,t}+u_{2n-1,t}}{4} + \frac{b}{\varDelta\hspace{1pt}a} \sin \left( \frac{u_{2n+1}-u_{2n}}{2} \right) 
	+\varDelta \hspace{1pt}ad \sin \left( \frac{u_{2n+1}+u_{2n}}{2} \right)
\nonumber \\[1mm]
&  - \frac{d}{\varDelta}  \sin \left( \frac{u_{2n}-u_{2n-1}}{2} \right) 
	+\varDelta \hspace{1pt}b  \sin \left( \frac{u_{2n}+u_{2n-1}}{2} \right)=0;
\label{5.13.1}
\end{align}
by eliminating $f_{2n-1}$ from (\ref{5.11.2}) and (\ref{5.12.1}), we have  
\begin{align}
& \frac{u_{2n,t}+u_{2n-2,t}}{4} - \frac{d}{\varDelta}  \sin \left( \frac{u_{2n}-u_{2n-1}}{2} \right) 
	-\varDelta \hspace{1pt}b  \sin \left( \frac{u_{2n}+u_{2n-1}}{2} \right)
\nonumber \\[1mm]
& +\frac{b}{\varDelta\hspace{1pt}a} \sin \left( \frac{u_{2n-1}-u_{2n-2}}{2} \right) 
	-\varDelta \hspace{1pt}ad \sin \left( \frac{u_{2n-1}+u_{2n-2}}{2} \right)=0.
\label{5.13.2}
\end{align}
\end{subequations}

By setting 
\mbox{$ad=b$}, 
(\ref{5.13}) can be 
expressed as
\begin{align}
& \frac{u_{2n+1,t}+u_{2n-1,t}}{2} + \frac{2d}{\varDelta} \left[ \sin \left( \frac{u_{2n+1}-u_{2n}}{2} \right) 
	 -\sin \left( \frac{u_{2n}-u_{2n-1}}{2} \right) \right]
\nonumber \\[1mm]
& 	+2\varDelta \hspace{1pt}b \left[ \sin \left( \frac{u_{2n+1}+u_{2n}}{2} \right)
	+\sin \left( \frac{u_{2n}+u_{2n-1}}{2} \right) \right]=0, 
\nonumber
\\[3mm]
& \frac{u_{2n,t}+u_{2n-2,t}}{2} - \frac{2d}{\varDelta} \left[ \sin \left( \frac{u_{2n}-u_{2n-1}}{2} \right) 
	-\sin \left( \frac{u_{2n-1}-u_{2n-2}}{2} \right) \right]
\nonumber \\[1mm]
& 	-2\varDelta \hspace{1pt}b \left[ \sin \left( \frac{u_{2n}+u_{2n-1}}{2} \right)
	+\sin \left( \frac{u_{2n-1}+u_{2n-2}}{2} \right)\right]=0,
\nonumber 
\end{align}
which 
are equivalent to 
%
\begin{subequations}
\label{5.15}
\begin{align}
& \frac{u_{2n+1,t}+u_{2n-1,t}}{2} + \frac{4d}{\varDelta} \sin \left( \frac{u_{2n+1}+u_{2n-1}-2u_{2n}}{4} \right) 
	 \cos \left( \frac{u_{2n+1}-u_{2n-1}}{4} \right) 
\nonumber \\[1mm]
& 	+4\varDelta \hspace{1pt}b \sin \left( \frac{u_{2n+1}+2u_{2n}+u_{2n-1}}{4} \right)
	\cos \left( \frac{u_{2n+1}-u_{2n-1}}{4} \right) =0, 
\label{5.15.1}
\\[5mm]
& \frac{u_{2n,t}+u_{2n-2,t}}{2} -\frac{u_{2n+1,t}+u_{2n-1,t}}{4} - \frac{u_{2n-1,t}+u_{2n-3,t}}{4}
\nonumber \\[1mm]
& 
- \frac{2d}{\varDelta} \left[ \sin \left( \frac{u_{2n+1}-u_{2n-1}}{4} \right) 
	 \cos \left( \frac{u_{2n+1}-2u_{2n}+u_{2n-1}}{4} \right) 
\right.
\nonumber \\[1mm]
& \hspace{11mm} \left. \mbox{}-\sin \left( \frac{u_{2n-1}-u_{2n-3}}{4} \right) 
	\cos \left( \frac{u_{2n-1}-2u_{2n-2}+u_{2n-3}}{4} \right)\right]
\nonumber \\[1mm]
&	-\varDelta \hspace{1pt}b \left[ \sin \left( \frac{u_{2n+1}+u_{2n}}{2} \right) 
 	 +3\sin \left( \frac{u_{2n}+u_{2n-1}}{2} \right) \right.
\nonumber \\[1mm]
& \hspace{11mm} \left. \mbox{}+ 3\sin \left( \frac{u_{2n-1}+u_{2n-2}}{2} \right)
	+\sin \left( \frac{u_{2n-2}+u_{2n-3}}{2} \right) \right]=0.
\label{5.15.2}
\end{align}
\end{subequations}

By further setting \mbox{$b=\frac{1}{16}$}, \mbox{$d=-1$} 
(and thus \mbox{$a=-\frac{1}{16}$}) and 
changing the 
dependent variable $u_{2n}$ to a new dependent variable $v_{2n}$ 
through the relation: 
\begin{equation}
u_{2n} =: \frac{u_{2n+1}+u_{2n-1}-\varDelta v_{2n}}{2}, 
\nonumber
\end{equation}
(\ref{5.15}) can be 
rewritten 
as a two-component system for $u_{2n-1}$ and $v_{2n}$: 
\begin{subequations}
\label{5.16}
\begin{align}
& \frac{u_{2n+1,t}+u_{2n-1,t}}{2} = \frac{4}{\varDelta} \sin \left( \frac{\varDelta v_{2n}}{4} \right) 
	 \cos \left( \frac{u_{2n+1}-u_{2n-1}}{4} \right) 
\nonumber \\[1mm]
& 	\hphantom{\frac{u_{2n+1,t}+u_{2n-1,t}}{2} =}
	-\frac{\varDelta}{4} \sin \left( \frac{2u_{2n+1}+2u_{2n-1}-\varDelta v_{2n}}{4} \right)
	\cos \left( \frac{u_{2n+1}-u_{2n-1}}{4} \right), 
\label{5.16.1}
\\[2mm]
&  \frac{v_{2n,t}+v_{2n-2,t}}{2}
\nonumber \\[1mm]
& 
= \frac{4}{\varDelta^2} \left[ \sin \left( \frac{u_{2n+1}-u_{2n-1}}{4} \right) 
	 \cos \left( \frac{\varDelta v_{2n}}{4} \right) 
	-\sin \left( \frac{u_{2n-1}-u_{2n-3}}{4} \right) 
	\cos \left( \frac{\varDelta v_{2n-2}}{4} \right)\right]
\nonumber \\[1mm]
&	\hphantom{=}\, -\frac{1}{8}\left[ \sin \left( \frac{3u_{2n+1}+u_{2n-1}-\varDelta v_{2n}}{4} \right) 
 	 +3\sin \left( \frac{u_{2n+1}+3u_{2n-1}-\varDelta v_{2n}}{4} \right) \right.
\nonumber \\[1mm]
& \hphantom{=} \hspace{9mm} \left. \mbox{}+ 3\sin \left( \frac{3u_{2n-1}+u_{2n-3}-\varDelta v_{2n-2}}{4} \right)
	+\sin \left( \frac{u_{2n-1}+3u_{2n-3}-\varDelta v_{2n-2}}{4} \right) \right].
\label{5.16.2}
\end{align}
\end{subequations}

The Lax pair for 
(\ref{5.16}) is given by 
%
\begin{align}
& L_{2n-1} (\lambda) = \left[
\begin{array}{cc}
 1 & 0 \\
 0 & 1 \\
\end{array}
\right] - 
\frac{\varDelta}{\lambda} \left[
\begin{array}{cc}
 \cos \left(\frac{u_{2n+1}+3u_{2n-1}-\varDelta v_{2n}}{4} \right) & \sin \left( \frac{u_{2n+1}+3u_{2n-1}-\varDelta v_{2n}}{4} \right) 
\\[3mm]
 \sin \left( \frac{u_{2n+1}+3u_{2n-1}-\varDelta v_{2n}}{4} \right) & -\cos \left( \frac{u_{2n+1}+3u_{2n-1}-\varDelta v_{2n}}{4} \right) \\
\end{array}
\right],
\nonumber 
\\[2mm]
& V_{2n-1} (\lambda)
= \frac{\lambda}{16} \left[
\begin{array}{cc}
 -1 & 0 \\
 0 & 1 \\
\end{array}
\right] + \left[
\begin{array}{cc}
 0 & f_{2n-1} \\
 -f_{2n-1}  & 0 \\
\end{array}
\right] 
+\frac{1}{\lambda} \left[
\begin{array}{cc}
 \cos u_{2n-1} & \sin u_{2n-1} \\
 \sin u_{2n-1} & -\cos u_{2n-1} \\
\end{array}
\right],
\nonumber 
\\[2mm]
& L_{2n} (\lambda) = \frac{\lambda}{16} \left[
\begin{array}{cc}
 -\varDelta & 0 \\
 0 & \varDelta \\
\end{array}
\right] + \left[
\begin{array}{cc}
 \cos \left(\frac{u_{2n+1}-u_{2n-1}+\varDelta v_{2n}}{4} \right) & -\sin \left( \frac{u_{2n+1}-u_{2n-1}+\varDelta v_{2n}}{4} \right) \\[3mm]
 \sin \left( \frac{u_{2n+1}-u_{2n-1}+\varDelta v_{2n}}{4} \right) & \cos \left( \frac{u_{2n+1}-u_{2n-1}+\varDelta v_{2n}}{4} \right) \\
\end{array}
\right],
\nonumber 
\\[2mm]
& V_{2n} (\lambda)
= \frac{\lambda}{16} \left[
\begin{array}{cc}
 -1 & 0 \\
 0 & 1 \\
\end{array}
\right] + \left[
\begin{array}{cc}
 0 & f_{2n} \\
 -f_{2n}  & 0 \\
\end{array}
\right] 
\nonumber \\[2mm]
& \hphantom{V_{2n} (\lambda)=} 
+\frac{1}{\lambda} \left[
\begin{array}{cc}
 \cos \left( \frac{u_{2n+1}+u_{2n-1}-\varDelta v_{2n}}{2} \right) 
	& \sin \left( \frac{u_{2n+1}+u_{2n-1}-\varDelta v_{2n}}{2} \right)\\[3mm]
 \sin \left( \frac{u_{2n+1}+u_{2n-1}-\varDelta v_{2n}}{2} \right) 
	& -\cos \left( \frac{u_{2n+1}+u_{2n-1}-\varDelta v_{2n}}{2} \right) \\
\end{array}
\right],
\nonumber 
\end{align}
where
\begin{align}
& f_{2n-1} =- \frac{u_{2n-1,t}-u_{2n-3,t}+\varDelta v_{2n-2,t}}{8} 
	- \frac{1}{\varDelta} \sin \left( \frac{u_{2n-1}-u_{2n-3}+\varDelta v_{2n-2}}{4} \right) 
\nonumber \\[1mm]
& \hphantom{f_{2n-1} =}
	-\frac{\varDelta}{16} \sin \left( \frac{3u_{2n-1}+u_{2n-3}-\varDelta v_{2n-2}}{4} \right),
\nonumber 
\\[2mm]
& f_{2n} = \frac{u_{2n+1,t}-u_{2n-1,t}+\varDelta v_{2n,t}}{8} - \frac{1}{\varDelta} \sin \left( \frac{u_{2n+1}-u_{2n-1}+\varDelta v_{2n}}{4} \right) 
\nonumber \\[1mm]
& \hphantom{f_{2n} =}
	+\frac{\varDelta}{16}\sin \left( \frac{3u_{2n+1}+u_{2n-1}-\varDelta v_{2n}}{4} \right).
\nonumber 
\end{align}
In 
the continuous 
limit \mbox{$\varDelta \to 0$}, 
(\ref{5.16})
reduces to 
(\ref{sG_2c}); 
this becomes 
conspicuous if we relabel the lattice sites 
as 
\mbox{$u_{2n-3} \to u_{n-1}$}, \mbox{$u_{2n-1} \to u_{n}$}, \mbox{$u_{2n+1} \to u_{n+1}$}, 
\mbox{$v_{2n-2} \to v_{n-1}$}, \mbox{$v_{2n} \to v_{n}$} 
and 
interpret 
$\varDelta$ as the lattice 
spacing. 
Thus, (\ref{5.16}) 
provides an integrable space discretization of the 
system (\ref{sG_2c}).

\section{Discretization 
of the sine-Gordon equation \mbox{$u_{tt} - u_{xx} + \sin u =0$}}

As a space-discrete 
analog of the Lax pair (\ref{TF_Lax4}) (with \mbox{$a=\frac{1}{16}$}) 
for the sine-Gordon equation in laboratory coordinates (\ref{sG_light2}), 
we consider the following 
Lax pair: 
\begin{subequations}
\label{sd_Lax6}
\begin{align}
& L_n (\lambda) = \left[
\begin{array}{cc}
 1 & 0 \\
 0 & 1 \\
\end{array}
\right] 
\nonumber \\[1mm]
& \mbox{}+ 
\varDelta \left[
\begin{array}{cc}
 -\frac{\lambda}{16} + Y_n - \frac{1}{\lambda} \cos \left( \frac{u_{n+1}+u_n}{2} \right) 
	& \frac{q_{n+1} + q_{n}}{2} - \frac{1}{\lambda} \sin \left( \frac{u_{n+1}+u_n}{2} \right) \\[1mm]
 -\frac{q_{n+1} + q_{n}}{2} - \frac{1}{\lambda} \sin \left( \frac{u_{n+1}+u_n}{2} \right) 
	& \frac{\lambda}{16} + Y_n + \frac{1}{\lambda} \cos \left( \frac{u_{n+1}+u_n}{2} \right) \\
\end{array}
\right],
\label{L_n6.1}
\\[2mm]
& V_n (\lambda)
= \frac{\lambda}{16} \left[
\begin{array}{cc}
 -1 & 0 \\
 0 & 1\\
\end{array}
\right] + \left[
\begin{array}{cc}
 0 & q_n \\
 -q_n  & 0 \\
\end{array}
\right]
+\frac{1}{\lambda} \left[
\begin{array}{cc}
 \cos u_n & \sin u_n \\
 \sin u_n & -\cos u_n \\
\end{array}
\right], 
\label{V_n6.1}
\end{align}
\end{subequations}
where 
$\varDelta$ is a (typically small but nonzero) lattice parameter 
and 
$Y_n$ and $q_n$ are some functions of the 
dependent variable $u_n$ 
to be determined. 

Substituting the Lax pair (\ref{sd_Lax6}) into the semi-discrete zero-curvature equation (\ref{sdLax_eq00}), 
we obtain 
\begin{subequations}
\label{system_6.1}
\begin{align}
& \varDelta \frac{u_{n+1,t}+u_{n,t}}{2} +2 \left( 1+\varDelta \hspace{1pt} Y_n \right) \sin \left( \frac{u_{n+1}-u_n}{2} \right) 
\nonumber \\[1mm] 
& +  \varDelta \left( q_{n+1} + q_n \right) \left[ 1+ \cos \left( \frac{u_{n+1}-u_n}{2} \right) \right] 
=0,
\label{6.1.1}
\\[2mm]
& Y_{n,t} +\frac{1}{16} \left( \cos u_{n+1} - \cos u_n \right) + \frac{q_{n+1}^2-q_n^2}{2} =0,
\label{6.1.2}
\\[2mm]
& \varDelta \hspace{1pt} \frac{q_{n+1,t}+q_{n,t}}{2} - \left( 1+\varDelta \hspace{1pt}Y_n \right) \left( q_{n+1} -q_n \right) 
	-\frac{\varDelta}{16} \left( \sin u_{n+1} + \sin u_n \right) 
\nonumber \\[1mm]
& - \frac{\varDelta}{8} \sin \left( \frac{u_{n+1}+u_n}{2} \right)=0.  
\label{6.1.3}
\end{align}
\end{subequations}
Note that (\ref{6.1.1}) can be rewritten as 
\begin{align}
& \varDelta \hspace{1pt} \frac{u_{n+1,t}+u_{n,t}}{8\cos^2 \left( \frac{u_{n+1}-u_n}{4} \right)} 
	+ \left( 1+\varDelta \hspace{1pt} Y_n \right)  \tan \left( \frac{u_{n+1}-u_n}{4} \right)
+ \varDelta \hspace{1pt} \frac{q_{n+1}+q_n}{2} =0.
\label{6.1.4}
\end{align}

The system (\ref{system_6.1}) 
implies the relation:
\begin{equation}
\left[ \varDelta^2 \frac{\left( q_{n+1}+q_{n} \right)^2}{4}  + \left( 1+ \varDelta \hspace{1pt}Y_n \right)^2 
	- \frac{\varDelta^2}{8} \cos \left( \frac{u_{n+1}+u_n}{2} \right)
\right]_t = 0,
\nonumber
\end{equation}
which
can also be 
obtained as a direct consequence of 
(\ref{sd-cons}). 
We set 
\begin{equation}
 \varDelta^2 \frac{\left( q_{n+1}+q_{n} \right)^2}{4}  + \left( 1+ \varDelta \hspace{1pt}Y_n \right)^2 
	- \frac{\varDelta^2}{8} \cos \left( \frac{u_{n+1}+u_n}{2} \right) = 1. 
\label{cons6.1}
\end{equation}
Combining (\ref{6.1.4}) with (\ref{cons6.1}),
we obtain a quadratic equation for \mbox{$ 1+ \varDelta \hspace{1pt}Y_n$}: 
\begin{align}
& \left( 1+ \varDelta \hspace{1pt}Y_n \right)^2 
+ \varDelta \hspace{1pt} \frac{u_{n+1,t}+u_{n,t}}{4} \tan \left( \frac{u_{n+1}-u_n}{4} \right) 
	\left( 1+ \varDelta \hspace{1pt}Y_n \right)
+ \frac{\varDelta^2\left(u_{n+1,t}+u_{n,t}\right)^2}{64\cos^2 \left( \frac{u_{n+1}-u_n}{4} \right)} 
\nonumber \\[1mm]
& 
- \cos^2 \left( \frac{u_{n+1}-u_n}{4}\right) \left[ 1+\frac{\varDelta^2}{8} \cos \left( \frac{u_{n+1}+u_n}{2} \right)\right] =0, 
\nonumber
\end{align}
and 
choose the proper solution 
of this 
quadratic 
equation 
as 
\begin{align}
& 1+ \varDelta \hspace{1pt}Y_n =-\varDelta \hspace{1pt} \frac{u_{n+1,t}+u_{n,t}}{8} 
\tan \left( \frac{u_{n+1}-u_n}{4} \right) 
\nonumber \\[1mm]
& \mbox{} + \cos \left( \frac{u_{n+1}-u_n}{4}\right) \sqrt{
1+\frac{\varDelta^2}{8} \cos \left( \frac{u_{n+1}+u_n}{2} \right) 
- \frac{\varDelta^2\left( u_{n+1,t}+u_{n,t} \right)^2}{64\cos^2 \left(\frac{u_{n+1}-u_n}{4}\right) }},
\label{Y_exp}
\end{align}
where the square root function 
is defined as the Maclaurin series in $\varDelta^2$. 

From 
(\ref{6.1.4}) and (\ref{Y_exp}), we obtain an explicit expression for $\varDelta \hspace{1pt} \frac{q_{n+1}+q_n}{2}$: 
\begin{align}
& \varDelta\hspace{1pt} \frac{q_{n+1}+q_n}{2} =
-\varDelta \hspace{1pt} \frac{u_{n+1,t}+u_{n,t}}{8} 
\nonumber \\[1mm]
& 
-\sin \left( \frac{u_{n+1}-u_n}{4} \right) 
\sqrt{1+\frac{\varDelta^2}{8} \cos \left( \frac{u_{n+1}+u_n}{2} \right)
- \frac{\varDelta^2\left( u_{n+1,t}+u_{n,t} \right)^2}{64\cos^2 \left( \frac{u_{n+1}-u_n}{4}\right)}}.
\label{q_exp1}
\end{align}
Substituting (\ref{q_exp1}) into (\ref{6.1.3}), we obtain 
\begin{align}
&  \left( 1+\varDelta \hspace{1pt}Y_n \right) \left( q_{n+1} -q_n \right)  = 
-\varDelta \hspace{1pt} \frac{u_{n+1,tt}+u_{n,tt}}{8} 
\nonumber \\[1mm]
& -\left\{ \sin \left( \frac{u_{n+1}-u_n}{4}\right) \sqrt{
1+\frac{\varDelta^2}{8} \cos \left( \frac{u_{n+1}+u_n}{2} \right)
- \frac{\varDelta^2\left( u_{n+1,t}+u_{n,t} \right)^2}{64\cos^2 \left( \frac{u_{n+1}-u_n}{4}\right) }} \right\}_t
\nonumber \\[1mm]
& -\frac{\varDelta}{16} \left( \sin u_{n+1} + \sin u_n \right) 
- \frac{\varDelta}{8} \sin \left( \frac{u_{n+1}+u_n}{2} \right).  
\label{q_exp2}
\end{align}
%
Combining (\ref{q_exp1}) 
with (\ref{q_exp2}), we obtain 
\begin{align}
&  2q_{n+1} = 
-\varDelta \hspace{1pt} \frac{u_{n+1,tt}+u_{n,tt}}{8\left( 1+\varDelta \hspace{1pt}Y_n \right) } 
-\frac{1}{1+\varDelta \hspace{1pt}Y_n} 
\nonumber \\[1mm]
& \mbox{} \times \left\{ \sin\left( \frac{u_{n+1}-u_n}{4} \right) \sqrt{
1+\frac{\varDelta^2}{8} \cos \left( \frac{u_{n+1}+u_n}{2} \right)
- \frac{\varDelta^2\left( u_{n+1,t}+u_{n,t} \right)^2}{64\cos^2 \left( \frac{u_{n+1}-u_n}{4}\right) }} \right\}_t
\nonumber \\[1mm]
& -\varDelta \hspace{1pt} \frac{\sin u_{n+1} + \sin u_n}{16 \left( 1+\varDelta \hspace{1pt}Y_n \right)}
- \varDelta \hspace{1pt} \frac{\sin \left( \frac{u_{n+1}+u_n}{2} \right)}{8\left( 1+\varDelta \hspace{1pt}Y_n \right)}
-\frac{u_{n+1,t}+u_{n,t}}{4}
\nonumber \\[1mm]
& -\frac{2}{\varDelta}\sin \left( \frac{u_{n+1}-u_n}{4} \right) \sqrt{
1+\frac{\varDelta^2}{8} \cos \left( \frac{u_{n+1}+u_n}{2} \right) 
- \frac{\varDelta^2\left( u_{n+1,t}+u_{n,t} \right)^2}{64\cos^2 \left( \frac{u_{n+1}-u_n}{4}\right) }}, 
\label{q_exp3}
\end{align}
and 
\begin{align}
&  2q_{n} = 
\varDelta \hspace{1pt} \frac{u_{n+1,tt}+u_{n,tt}}{8\left( 1+\varDelta \hspace{1pt}Y_n \right) } 
+\frac{1}{1+\varDelta \hspace{1pt}Y_n}
\nonumber \\[1mm]
& \mbox{} \times \left\{ \sin \left( \frac{u_{n+1}-u_n}{4} \right) \sqrt{
1+\frac{\varDelta^2}{8} \cos \left( \frac{u_{n+1}+u_n}{2} \right) 
- \frac{\varDelta^2\left( u_{n+1,t}+u_{n,t} \right)^2}{64\cos^2 \left( \frac{u_{n+1}-u_n}{4}\right) }} \right\}_t
\nonumber \\[1mm]
& +\varDelta \hspace{1pt} \frac{\sin u_{n+1} + \sin u_n}{16\left( 1+\varDelta \hspace{1pt}Y_n \right)}
+\varDelta \hspace{1pt} \frac{\sin \left( \frac{u_{n+1}+u_n}{2} \right)}{8\left( 1+\varDelta \hspace{1pt}Y_n \right)}
-\frac{u_{n+1,t}+u_{n,t}}{4}
\nonumber \\[1mm]
& -\frac{2}{\varDelta}\sin \left( \frac{u_{n+1}-u_n}{4} \right) \sqrt{
1+\frac{\varDelta^2}{8} \cos \left( \frac{u_{n+1}+u_n}{2} \right) 
- \frac{\varDelta^2\left( u_{n+1,t}+u_{n,t} \right)^2}{64\cos^2 \left( \frac{u_{n+1}-u_n}{4}\right) }}.
\label{q_exp4}
\end{align}
Finally, comparing (\ref{q_exp3}) with (\ref{q_exp4}), we arrive at the equation of motion for the single dependent variable 
$u_n$: 
\begin{align}
& \frac{u_{n+1,tt}+u_{n,tt}}{4\left( 1+\varDelta \hspace{1pt}Y_n \right) } 
+\frac{u_{n,tt}+u_{n-1,tt}}{4\left( 1+\varDelta \hspace{1pt}Y_{n-1} \right) } -\frac{u_{n+1,t} - u_{n-1,t}}{2\varDelta}
\nonumber \\[1mm]
& +\frac{2}{\varDelta \left( 1+\varDelta \hspace{1pt}Y_n \right)} 
\nonumber \\[1mm]
&
\times \left\{ \sin \left( \frac{u_{n+1}-u_n}{4} \right) \sqrt{
1+\frac{\varDelta^2}{8} \cos \left( \frac{u_{n+1}+u_n}{2} \right) 
- \frac{\varDelta^2\left( u_{n+1,t}+u_{n,t} \right)^2}{64\cos^2 \left( \frac{u_{n+1}-u_n}{4}\right) }} \right\}_t
\nonumber \\[1mm]
& 
+\frac{2}{\varDelta \left( 1+\varDelta \hspace{1pt}Y_{n-1} \right)} 
\nonumber \\[1mm]
& \times \left\{ \sin \left( \frac{u_{n}-u_{n-1}}{4} \right) \sqrt{
1+\frac{\varDelta^2}{8} \cos \left( \frac{u_{n}+u_{n-1}}{2} \right) 
- \frac{\varDelta^2\left( u_{n,t}+u_{n-1,t} \right)^2}{64\cos^2 \left( \frac{u_{n}-u_{n-1}}{4}\right) }} \right\}_t
\nonumber \\[1mm]
& 
-\frac{4}{\varDelta^2}\sin \left( \frac{u_{n+1}-u_n}{4} \right) 
\sqrt{1+\frac{\varDelta^2}{8} \cos \left( \frac{u_{n+1}+u_n}{2} \right) 
- \frac{\varDelta^2\left( u_{n+1,t}+u_{n,t} \right)^2}{64\cos^2 \left( \frac{u_{n+1}-u_n}{4}\right)}}
\nonumber \\[1mm]
& +\frac{4}{\varDelta^2}\sin \left( \frac{u_{n}-u_{n-1}}{4} \right) 
\sqrt{1+\frac{\varDelta^2}{8} \cos \left( \frac{u_{n}+u_{n-1}}{2} \right) 
- \frac{\varDelta^2\left( u_{n,t}+u_{n-1,t} \right)^2}{64\cos^2 \left( \frac{u_{n}-u_{n-1}}{4}\right)}}
\nonumber \\[1mm]
& +\frac{ \sin u_{n+1} + \sin u_n +2\sin \left( \frac{u_{n+1}+u_n}{2} \right)}{8\left( 1+\varDelta \hspace{1pt}Y_n \right)}
+ \frac{\sin u_{n} + \sin u_{n-1}+2\sin \left( \frac{u_{n}+u_{n-1}}{2} \right)}{8\left( 1+\varDelta \hspace{1pt}Y_{n-1}\right)}=0,
\label{u_eqn}
\end{align}
where \mbox{$ 1+ \varDelta \hspace{1pt}Y_n$} is expressed as 
(\ref{Y_exp}). 
The Lax pair for (\ref{u_eqn}) 
is 
given by 
(\ref{sd_Lax6}), where 
$\varDelta \hspace{1pt} \frac{q_{n+1}+q_n}{2}$ in (\ref{L_n6.1}) 
is expressed as 
(\ref{q_exp1}) and 
$q_n$ in (\ref{V_n6.1})
is expressed as 
(\ref{q_exp3}) with \mbox{$n \to n-1$} divided by two. 
By 
taking the continuous limit \mbox{$\varDelta \to 0$}, 
(\ref{u_eqn}) 
indeed 
reduces to the sine-Gordon equation in laboratory coordinates (\ref{sG_light2}).

\section{Conclusions} 

In this paper, we addressed the problem of 
constructing 
integrable 
semi-discretizations of the sine-Gordon equation 
in non-characteristic coordinates. 
We considered a space discretization of the 
zero-curvature representation 
in three distinct cases of non-characteristic coordinates, 
while 
assuming that 
the 
functional form 
of the temporal Lax matrix 
with respect to the spectral parameter 
remains 
the same. 
This 
guarantees that 
%
the semi-discrete Lax pairs constructed in this paper are not fake; 
each semi-discrete Lax pair allows to derive an infinite number 
of conservation laws and to apply the inverse scattering transform~\cite{
TF74,ZTF75,Kaup75}
after 
swapping the roles of the spatial and temporal variables conceptually. 

Thus, we
obtained the integrable 
space-discrete 
sine-Gordon equations (\ref{sdsG2}), (\ref{4.8}) and (\ref{u_eqn}) 
with (\ref{Y_exp}). 
However, for the most interesting case of laboratory coordinates, 
the integrable 
space discretization 
(\ref{u_eqn}) with (\ref{Y_exp}) 
is not 
aesthetically pleasing. 
As a remedy, 
we 
rewrote the sine-Gordon equation in laboratory coordinates (\ref{sG_light2}) as 
the two-component evolutionary system (\ref{sG_2c})
and proposed 
the 
integrable 
space discretization (\ref{5.16}). 
Another integrable 
space 
discretization of 
the same 
system, known as the lattice sine-Gordon model, 
was proposed by 
Izergin and Korepin~\cite{IzeKor81} in the quantum setting 
(see~\cite{IzeKor86,Tarasov86,BogoIzeKor93}
for the classical 
lattice sine-Gordon model). 
The lattice sine-Gordon model 
is 
conceptually elegant from the point of view of the Hamiltonian 
formalism 
and the $r$-matrix structure; however, 
it is not 
so easy 
to understand intuitively
that the lattice sine-Gordon model
reduces to 
the two-component 
system (\ref{sG_2c}) in the continuous limit. 

%

\end{document}